\documentclass [12pt] {article}

\catcode`\@=11
\@addtoreset{footnote}{section}
\catcode`\@=12

\newcommand{\up}{\uparrow}
\newcommand{\dn}{\downarrow}

\newcommand{\msf}{\mathsf}

\newcommand {\myoI} [1] {\oint \! #1 \,}

\newcommand{\atvalue}[2]{{\left. #1 \:\right|^{#2}}}

\newcommand{\srd}{{\mathbf f}}

\newcommand{\srl}{{\mathbf g}}

\newcommand{\sbr}{{\mathbf k}}

\usepackage{amssymb}
\usepackage {latexsym}

\newcommand {\pa} {\partial}

\newcommand {\id} {\mathbb I}

\newcommand {\beq} {\begin {equation}}
\newcommand {\eeq} {\end {equation}}

\newcommand {\beqn} {\begin {displaymath}}
\newcommand {\eeqn} {\end {displaymath}}

\newcommand {\beqar} {\begin {eqnarray}}
\newcommand {\eeqar} {\end {eqnarray}}

\newcommand {\beqarn} {\begin {eqnarray*}}
\newcommand {\eeqarn} {\end {eqnarray*}}

\newcommand {\nono} {\nonumber \\ {}}

\newcommand {\bary} {\begin {array}}
\newcommand {\eary} {\end {array}}

\renewcommand {\cal}[1] {\mathcal {#1}}

\newcommand {\half} {\frac 1 2}

\newcommand {\csp} {\;\;}

\newcommand {\eqr} [1]  {{(eq. \ref {eq:#1})}}
\newcommand {\Eqr} [1]  {{(Eq. \ref {eq:#1})}}

\newcommand {\secr} [1] {\S \ref{sec:#1}}

\newcommand {\ignoretext} [1] {}

\newcommand {\ZZ} {\mathbb Z}

\newcommand {\RR} {{\mathbb  R}}

\newcommand {\CC} {{\mathbb{C}}}

\newcommand {\BSk}[1] [{ }] {\left|\left.#1\right>\!\right)}
\newcommand {\CSk} [1] [{ }] {\left|\left.#1\right>\!\right>}

\newcommand{\brac}[1]{\left\{{#1}\right\}}

\newcommand{\brak}[1]{\left[{#1}\right]}

\makeatletter
\@addtoreset{equation}{section}

\usepackage{hyperref}

\begin{document}

\begin{titlepage}
\hfill
\vbox{ 
\halign{#\hfil     \cr   
           CERN-TH/2002-318 \cr
           hep-th/0212338  \cr
           } 
      }  
\vspace*{25mm}
\begin{center}
{\Large {\bf  From Boundaries To Conditions \\
                Over Superspace}\\}
\vspace*{15mm}
{
Zheng Yin}

\vspace*{5mm}
{\it {Theory Division, CERN \\
CH-1211 Geneva  23, Switzerland\\
{\footnotesize Zheng.Yin@cern.ch}}}\\

\vspace{10mm}

\end{center}

\begin{abstract}
$N=1$ and $2$ superconformal boundary conditions are shown to be the
consequence of a boundary on the worldsheet superspace with positive
codimension in the anticommuting subspace.  In addition to the 
well-known boundary conditions, I also find two new infinite series of 
$N=2$ boundary states.  Their free field realizations are given.  
A self-contained development of 2d superspace leads to
new perspectives on this subject.
\end{abstract}
\end{titlepage}


\section {Introduction}

Recently there has been some reexamination of 2d superconformal field
theory with boundaries \cite{Albertsson:2001dv,Lindstrom:2002jb},
elaborating the superconformal boundary conditions frequently used
\cite{ Callan:1988st,Ooguri:1996ck} for the study of D-branes
\cite{Polchinski:1996na}.  The setting was non-linear $\sigma$-models
with $N=1$ or $N=2$ worldsheet superconformal symmetry.  The analysis
was versed in the language of the target space: fields
represent coordinates on the target space or sections of vector bundles
over it, and coupling constants encodes its geometric feature.  This is
appropriate for relating the boundary conditions to the geometry of
D-branes \cite{Ooguri:1996ck,Albertsson:2002qc}.

The worldsheet origin of boundary suggests an orthogonal viewpoint. 
Boundary conditions that have to be imposed reflect the modification
made to the original theory to consistently accommodate an abrupt end to
the space on which it is defined.  A fundamental and universal
classification of boundary conditions ought to be made solely in
worldsheet language without
being tied to a particular action formalism.  Boundary conformal field
theory starts with a choice of boundary conditions imposed between the
left and right moving chiral fields.  This is most precisely stated in
the boundary state formalism.  The only constraint seemed to be
consistency with the chiral algebra, so they are 
classified by their outer automorphism group.  Those 
differing by an inner automorphism are mapped into each by symmetries 
generated by the chiral currents themselves.   For a particular type of
actions such as non-linear $\sigma$-model, one can make
contact with the target space point of view by substituting the concrete
realization of the chiral fields into the abstract boundary conditions.

However, the above procedure still does not contain a derivation
of the boundary condition itself as the necessary consequence of a 
boundary.  If one doubts the need for such an understanding, consider
the following outer automorphism of the Virasoro algebra:\footnote
    {I am grateful to V.~Kac for pointing out this automorphism.}
\beq
    L_m \to -L_{-m}
\eeq
There are two problems with using this automorphism to define a boundary
condition, such as 
\beq
    (L_m + \tilde L_m) \BSk = 0
\eeq
for a boundary state $\BSk$.
First, algebraic consistency dictates that this automorphism must also
flip the sign of the central charge.  This is not permissible because
in a CFT the central charge is unlike any other element of the Lie
algebra --- it is just a number, and it has to be the same for the left
and right movers.  Another problem, which persists even when the central
charge vanishes, is that this boundary condition is not local.  On a
space-like boundary parameterized by $\sigma\in[0,2\pi)$, It relates the
left and right moving stress tensors by
\beq
    T(\sigma) + \tilde T(-\sigma)
\eeq
As we shall see in \secr{NIsZero}, this does not follow from the
presence of a boundary.

We can eliminate this problematic automorphism by considering only those
leaving the central charge invariant and further require that they
have an appropriate local origin in terms of the stress tensor.
This makes clear that a derivation of boundary
conditions from worldsheet geometry would be desirable.  Unfortunately,
it is not clear how this could be done for general chiral algebras. 
Unlike the Virasoro algebra, they seem to represent
``internal'' degree of freedom mostly independent of the geometry of the
worldsheet.  However, as will be shown in the paper, boundary conditions
for $N=1$ and $N=2$ superconformal algebra (SCA) do have purely
worldsheet interpretation, if one is willing to complement the ordinary
worldsheet (the even part) with an \emph {odd} part parameterized by
Grassmann ``coordinates.''

Central to this paper is therefore the use of superspace to describe the
action of superconformal symmetries.  Worldsheet superspace has long
been used in string theory \cite{Fairlie:1973,Montonen:1974}, often
for writing down Lagrangians with manifest rigid supersymmetry
in terms of superfields \cite{Howe:1977xz,Martinec:1983um}.
Due to the different motivation and purpose of the present work, 
2d $N=1$ and $2$ superspace is developed ab initio in
this paper without reference to any Lagrangian or fundamental field.  As
a consequence issues such as auxiliary fields and off-shell closure is
not relevant.  Here the Grassmann coordinates are introduced solely as
organizational labels for the representations of the singular operator
production expansion (sOPE) algebra of the superconformal currents.  How they
transform under superconformal symmetry is then inferred from the
algebra, i.e. superspace emerges as a \emph{derived} construct.  This
philosophy differs significantly from that underlying previous works that aimed 
to combine superspace and superconformal symmetry, where the starting point had been the definition of a superdifferential.  For $N=1$
case, the resulting superspace agrees with \cite{Friedan:1986ge}.  For
$N=2$ the present results look superficially different from 
\cite{DiVecchia:1985iy,Cohn:1987wn,Kiritsis:1987np,Schoutens:1988ig}.
Both the compact and noncompact versions of $N=2$ SCA make appearance
here, classified and differentiated by nonequivalent reality conditions
of the superspace.  This has not been noted before, but after a field
redefinition the compact case does agree with those earlier results on
the transformation properties of superconformal tensors.

The technical advantage of this superspace approach is that it allows a
systematic and geometric derivation for the boundary conditions.  The analysis of pure 
conformal symmetry can be applied to the superconformal symmetries after
the appropriate adaptation.  At its most essential, a boundary
occupies an one dimensional curve on the even part of the superworldsheet.
 Those conformal symmetries that perturb it in the transverse direction
are broken \cite{ Cardy:1989ir}.  If the boundary also occupies the whole odd
part of the superworldsheet, all supersymmetries are broken.  The more
interesting case is when the boundary has finite codimension in the odd
part as well.  It shall be shown in this work that such superboundary
can be consistently defined and classified geometrically.  They preserve
half of the worldsheet supersymmetries.  The corresponding boundary
conditions for the SCA can also be inferred from geometry, modulo
central terms.  The latter can then be fixed easily by algebraic
arguments.  For $N=0$ and $1$ the standard boundary conditions are
recovered.  For $N=2$ SCA,
in addition to the well-known A and B types of boundary conditions I
also find two infinite classes of new ones.  They appear for closed
boundaries (e.g. boundary state or
one-loop diagram of open string) and have to do with the topology of the
superworldsheet.

Although this paper is concerned with boundary on the string worldsheet
and more generally 2d (super)conformal field theory, the idea of using
superspace geometry to to classify and derive boundary conditions can be
applied to other cases.  Boundaries and spacetime defects in
supergravity theory and the target space of superstring have recently
become an area of intensive research.  It would be interesting to see
whether the types of analysis presented here would be applicable to 
them.

The paper is organized as follows.  In section 1,the boundary condition
of Virasoro algebra is derived from the boundary geometry as a warm-up
for the superconformal case.  In section 2 $N=1$ superconformal
symmetry is considered.  In section 3 I proceed to $N=2$ superconformal
symmetry and find novel boundary states arising from nontrivial
superworldsheet topology.  In section 4 the novel boundary states are
realized in a free theory.  While this
paper is worldsheet oriented, these concrete solutions 
can be related to and used for D-branes in $N=2$ compactification
\cite{Ooguri:1996ck} and $N=2$ string theories
\cite{Cheung:2002yw,oy:2002N2DbraneToAppear2003}.

\section {Boundary Conditions for $N=0$}
\label{sec:NIsZero}

What follows can be thought as an elaboration of the argument in section
1 of \cite{Cardy:1989ir}.  It is intended as a warm up drill in which
the essential steps needed later for the superconformal cases are run
through in a context with little subtlety.

\paragraph{Worldsheet}
The 2d worldsheet is parametrized by coordinates $z$ and $\tilde z$.  
The field theory has conformal symmetry, realized projectively 
on the Hilbert space of the theory because of the 
conformal anomaly.  Denote as usual $T$ and $\tilde T$ the decoupled
left and right moving stress tensor.  The Lie algebra of conformal 
symmetry can be inferred from their sOPE \eqr{TTOPE}.
The operators of the theory are classified
into highest weight representations of T, each of which is characterized
by a conformal weight $h$ and a chiral primary field V satisfying the
following sOPE:
\beq
    T(z)\ldots V(w) = \frac {h V(w)} {(z-w)^2} 
    + \frac {\pa V(w)} {z-w}\ .
\eeq
Infinitesimal conformal transformations are parameterized by 
two functions $\epsilon(z)$ $\tilde
\epsilon(\tilde z)$ and the corresponding charge is\footnote{
    The convention for contour integration is counterclockwise for $z$
    and clockwise for $\tilde z$.}
\beq    \label{eq:GeneratorOfConformalTransformation}
\frac 1 {2\pi\imath} \left(\myoI{dz} \epsilon (z) T(z) 
 + \myoI{d\tilde z} \tilde \epsilon (\tilde z) \tilde T(\tilde z) \right)
\eeq
Therefore a chiral primary field $V$ with conformal weight $h$ with 
respect to $T$ varies as
\beq
    \delta V = \imath (\epsilon \pa V + h \pa \epsilon V)
\eeq
Recall that a tensor field ${\msf V}$ on a manifold, which may consist
of several components, is defined with respect to a group acting on that
manifold. It transform generically under an infinitesimal group element
$\epsilon$ by
\beq    \label{eq:GeneralVariationOfTensor}
    \delta_{\epsilon} {\msf V} = 
        \delta_{\epsilon} X^\mu \, \frac {\pa} {\pa{X^\mu}} {\msf V}
        + \Lambda_\epsilon^{[\Delta]} {\msf V}
\eeq
Here $\delta_{\epsilon}$ alone represents the action of the group on the
coordinate of the manifold and $\Lambda_{\epsilon}^{\Delta}$ represent
action of the group on the tensor.  $\Delta$ denote properties
that distinguish different tensor fields. Hence a chiral
primary field of weight $h$ with respect to $T$ is a tensor
field for conformal transformation with
\beq
    \delta_\epsilon z = \imath\epsilon(z)\ , \csp 
    \Lambda_\epsilon^{[h]} = \imath h \pa \epsilon(z)
\eeq
Therefore we recovers the spacetime action of a conformal transformation
\beq    \label {eq:ConformalTransformationOnz}
    \delta z = \imath\epsilon\ , \csp 
    \delta \tilde z = \imath\tilde \epsilon\ .
\eeq
Usually conformal transformation is simply \emph{defined} as such.  
The procedure here of deriving it from the sOPE of between
currents and tensor fields may seem circuitous, but it guarantees the
correct result once an sOPE algebra is given.  It is well suited for
the superconformal cases where geometric intuition is elusive.

\paragraph{Compatible reparameterization}
Given \eqr{ConformalTransformationOnz}, 
one can ask what kind of coordinate redefinition 
\beq
    z' = f(z, \tilde z)\ , \csp \tilde z' = f'(z, \tilde z)\ .
\eeq
leaves the form of the variation invariant, i.e. 
\beq
    \delta z' = \epsilon'(z)\ , \csp 
    \delta z' = \tilde \epsilon' (\tilde z)\ .
\eeq
for some $\epsilon'$.
This gives an automorphism of the sOPE
algebra realizable as a reparameterization of the worldsheet.
They are important for the classification of boundary conditions 
and, in conjunction with reality condition, superconformal field theories.
The solution for $f$ is clearly that $f$ should only depend on $z$, and
$f$ on $\tilde z$.  Hence they are ``almost'' conformal transformations
themselves.  To remove the qualifier ``almost'' we have to take into
account the reality condition.

\paragraph{Reality condition}
Up to this point I have not really specified the signature of the worldsheet
metric.  The previous equations are signature independent.  The effect
of the latter shows up in the \emph{reality condition}.  If the worldsheet is
Euclidean, $z$ and $\tilde z$ are complex and related by 
\beq
    z^* = \tilde z\ .
\eeq
For a Lorentzian worldsheet, $z$ and $\tilde z$ are both real and
unrelated.
For studying field theory in the operator formalism it is necessary that
the worldsheet is Lorentzian, and henceforth in this note only this case
will be considered.

Generally the reality condition on a Lorentzian worldsheet is 
$z^* = g(z)$
Under reparameterization $z' = f(z)$, $g$ transforms as 
\beq
    g(z) \to g'(z') = f^* \circ g \circ f^{(-1)} (z')
\eeq
where $f^{(-1)}$ denote the inverse function of $f$.  
In the literature,
one often uses the exponentiated and Wick-rotated notation in which $t$
is replaced by $\tau = \imath t$ and the coordinate $z =
\exp(\imath\sigma_L) = \exp{(\tau+\imath \sigma)}$, with $\sigma$ the
compactified spatial direction of the worldsheet.  The cylindrical 
worldsheet is mapped into a two punctured complex plane compactified at
infinity. To avoid confusion I shall continue to use $\tilde z$ to
denote the right moving coordinate and the superscript ${}^*$ to denote
conjugation in the Lorentzian sense with the appropriate analytic
continuation. The reality condition in this convention becomes 
\beq
    z^*= 1/z
\eeq
Unless specified otherwise, I shall follow this convention in the rest
of this work.  This is especially convenient for performing contour
integration with the sOPE's but introduces some complication for the
boundary and reality conditions.

Conformal transformation should of course respect the reality condition,
which leads to the reality conditions for $\epsilon$ and
$\tilde\epsilon$:
\beq    \label{eq:RealityConditionForepsilon}
    (\epsilon(z))^*= z^{-2}\epsilon(z)\ ,\csp 
    (\tilde \epsilon(\tilde z))^*= \tilde z^{-2}\tilde\epsilon(\tilde z)
\eeq
The reparameterization consider earlier should also leave the reality
condition invariant.  This leads to the reality condition on $f$.  It
still best to state it in terms of the unexponentiated, un-Wick-rotated
coordinates: $f(\sigma_L)$ and $\tilde f(\sigma_R)$ are both real.  In
this paper, 
reparameterizations 
compatible with infinitesimal (super)conformal transformations 
and a chosen reality condition are called 
\emph{reparameterization automorphisms}.  They obviously form a group.  
They are important for classifying boundary conditions.  It is also
clear that transition functions between different patches of the
(super)worldsheet must be elements of this group.
Note that in the present case they
are just the usual conformal transformations.  They are generated by
\eqr{GeneratorOfConformalTransformation} subjected to the appropriate
reality conditions on $\epsilon$ plus the Hermiticity of $T$ and 
$\tilde T$. It means that they are in fact inner automorphisms. For the
superconformal cases this will no longer be true.

\paragraph{Boundary condition}
Now consider a boundary on the Lorentzian worldsheet.  Locally
the boundary can be any one of three types: time-like, space-like, and null. 
The last type is located at a fixed value of say $z$ but have no
restriction locally for $\tilde z$.  It leads to intriguing 
pathologies and is rarely used in the literature.  Henceforth I 
shall require the boundary to be nowhere null.  
Thus they are globally either time-like or
space-like.  Analytically this can be seen as follows.  
A boundary condition that is nowhere null can be written as 
\beq
   \tilde z = k(z)\ .
\eeq
Then a reparameterization automorphism acts on $k$ by 
\beq
    k \to k' = \tilde f \circ k \circ f^{(-1)} 
\eeq
Equivalence classes of $k$ with respect to this 
classifies the boundary condition.  A time-like
boundary can always be brought to be the positive real axis.
\beq    \label{eq:PositiveRealAxis}
    \tilde z = z \geq 0
\eeq
and space-like boundary to 
\beq    \label{eq:UnitCircle}
    z \tilde z= 1\ .
\eeq
The conformal transformations leaving these equation 
invariant are respectively 
\beq    \label{eq:BoundaryConditionForCTTimelike}
    \atvalue{\epsilon = \tilde \epsilon}{z = \bar z}
\eeq
and 
\beq    \label{eq:BoundaryConditionForCTSpacelike}
    \atvalue{\bar z \epsilon = - z \tilde \epsilon}{|z| = 1}
\eeq

What does this imply for the currents and charges in the theory that
generates these symmetries?  Consider first an arbitrary symmetry of the
worldsheet that leaves invariant a boundary of the space on which the
theory is defined.For a local quantum field theory, features of the
space on which it is defined is immune from quantum fluctuation, so a
boundary and the symmetries preserving it should
manifest themselves up to quantum anomalies.  Furthermore,
quantum anomaly can be computed locally in the bulk.  
Therefore a boundary preserving symmetry
is realized in the theory if and only if the same symmetry exist in the
absence of the boundary, and with the same cocycle if the realization is
projective.

The manner in which it is realized depend on the type of the boundary.  
If the boundary is space-like, it
represents a state in the QFT,\footnote{
    It is not a state in the proper Hilbert space because it would
    generically have infinite norm, but this is not a relevant issue
    here.}
while if it is time-like, the boundary modifies the theory itself.
In the language of string theory, it is a boundary state in the close 
string theory versus an open string theory in its own right.  
Suppose the symmetry is generated by some charge
$Q$ constructed from some conserved current $j_\mu$ as usual by 
$Q = \int d^{D-1}x J_0$.  If the boundary is space-like
and represented by some boundary state $\BSk{}$, the latter should be
invariant under the charge:
\beq
    Q \BSk{} = 0
\eeq
If the boundary is time-like, the charge $Q$ remains a constant of 
motion.  It follows that the boundary contribution to $\dot Q$ vanishes: 
\beq
    \int_{\pa \Sigma} dA \cdot J  = 0
\eeq
This is a constraint on the open string theory stipulating 
that no net flux of $Q$ charge leak through the boundary.

Now apply this to conformal symmetries, where 
\beq
j_z = \epsilon T, \csp 
j_{\bar z} = \tilde \epsilon \tilde T\ .
\eeq
Substituting this with \eqr{BoundaryConditionForCTTimelike} in
\eqr{GeneratorOfConformalTransformation}, one finds after factoring out
the arbitrary $\epsilon$ that
\beq    \label{eq:BoundaryConditionForT}
    \atvalue{T - \tilde T = 0}{\sigma = 0}
\eeq
for the open string theory and 
\beq
    (T - \frac {{\tilde z}^2} {z^2} \tilde T)  
        \BSk = 0
\eeq
for the boundary state.
Since
\beq
    T = \sum_{n\in \ZZ} L_m z^{-m-2}, \csp 
    \tilde T = \sum_{n\in \ZZ} \tilde L_m \tilde z^{-m-2} \ .
\eeq
the last equation means 
\beq    \label{eq:BoundaryConditionForL}
    (L_m  - \tilde L_{-m} ) \BSk = 0 \ .
\eeq
To be precise, the above procedure gives only the correct boundary
condition modulo central terms.  It is easy to verify algebraically that
that no correction is needed in this case.  In fact these are the usual
boundary conditions for CFT.

Of course one can think of other boundary conditions.  A priori, the
only algebraic constraint is that the relation between the left and
right moving chiral fields are consistent with their algebra, therefore
boundary conditions are corresponds to 
automorphism of the symmetry algebra that 
preserves the central charge and appropriate Hermiticity
condition. For Virasoro algebra that just leaves the inner
automorphisms, 
those generated by $L_m$ themselves and corresponding to
conformal transformations. On the other hand, the above analysis has
shown that the boundary condition on the stress tensor, following
strictly the requirement of preserving the geometric boundary, is
\eqr{BoundaryConditionForT} or \eqr{BoundaryConditionForL} for time-like
or space-like boundaries, after the boundary is brought to canonical
forms by a conformal transformation.  Therefore the other possible
boundary conditions just correspond to boundaries that have not been
brought to one of the two canonical forms.  Had there been an
automorphisms that do not have some origin from geometry (e.g. a
boundary or a crosscap), their interpretation in boundary CFT would have 
been suspect.

When a bigger chiral algebra is known for a CFT, one should likewise
consider boundary conditions for them as well
\cite{Ishibashi:1989kg,Cardy:1989ir}.
However, these symmetries in general acts on ``internal'' degrees of
freedom rather than worldsheet.  In the language of
\eqr{GeneralVariationOfTensor}, they only shows up in
$\Lambda^{[\Delta]}_\epsilon$.  The method developed above relies on
$\delta_\epsilon X^\mu$ and thus cannot be used to give a geometric
derivation of their boundary
conditions.  However, for superconformal symmetries this is possible by
enlarging the notion of worldsheet, and for these cases the systematic
approach of this work displays its power.  It is amusing to note that
the very notion of superspace will be derived ab initio from the sOPE
algebra of the (super)currents rather than conjured out of the void. 
Even here one may take the view that the
worldsheet coordinate $z$ and $\tilde z$ was reconstructed from the
Virasoro algebra.

\section {Boundary Conditions for $N=1$}

\paragraph{$N=1$ superworldsheet}
The $N=1$ superconformal algebra is the simplest extension to Virasoro
algebra with an interpretation purely in terms of superworldsheet geometry.  
Consider the left moving part. The $N=1$ superconformal algebra (SCA)
complements $T$ with a superpartner $G$ and their sOPE's is given in     the appendix. From this one may determine the multiplet structure of
irreducible tensor fields and their sOPE's with $T$ and $G$.  There is
just one kind, again characterized by a conformal weight $h$ but now
consists of two components fields.  They can be combined into a
\emph{superfield} as $\msf V = V^\dn + \theta V^\up$ by introducing a
Grassmann variable $\theta$.  For the time being it is merely an 
nilpotent label.

$V^\dn$ and $V^\up$ are both
chiral primary fields of $T$ with conformal weight $h$ and $h+1/2$
respectively.  Their sOPE's with $G$ are
\beqar  \label{eq:NOneOPEWithV}
    G(z)\ldots V^\dn(w) &=& \frac {V^\up} {z-w}\ , \nono
    G(z)\ldots V^\up(w) &=& \frac {2h \pa V^\dn} {(z-w)^2} 
    + \frac {\pa V^\dn} {z-w}
\eeqar
Infinitesimal $N=1$ superconformal transformations are generated by 
\beq
    \frac 1 {2\pi\imath} \left(\myoI{dz} \epsilon (z) T(z) 
    + \myoI{dz} a(z) G(z) \right).
\eeq
where $a$ is the Grassmann valued parameters for supersymmetry.
By contour integration of the OPE \eqr{NOneOPEWithV}, one finds that 
\beqar
    \delta \msf V &=& \imath[\epsilon \pa 
        + \half \pa \epsilon \theta \pa_\theta 
        + h \pa \epsilon ] \msf V \nono
    &+& \imath[-a \theta \pa + a \pa_\theta - 2 h \pa a \theta] 
    \msf V\ .
\eeqar
From this I deduce the action of the superconformal transformation on 
the superspace coordinate:
\beq    \label{eq:NOneActionOnSuperspace}
    \delta z = \imath(\epsilon - a \theta), \csp 
    \delta \theta = \imath(\half \pa \epsilon \theta + a)\ ,
\eeq
and $\theta$ now emerges as a coordinate on a \emph{superworldsheet}.
One then does the same for the right movers, introducing $\tilde\theta$ 
in the process.  The superworldsheet is therefore make up of an even
part parameterized by $z$ and $\tilde z$ and an odd part parameterized
by $\theta$ and $\tilde\theta$.  To be precise, the latter is fibered
over the former.  The transition function are elements of the 
reparameterization automorphism discussed below.

\paragraph{Compatible reparameterization}
Consider the reparameterization
\beq
    z' = f(z), \csp \theta' = \srd(z) \theta
\eeq
that leaves \eqr{NOneActionOnSuperspace} invariant.  In fact I have 
already
used it to eliminate $\bar z$ dependence and mixing with $\tilde
\theta$.  Solving now for 
\beq
    \delta z' = \epsilon' - a' \theta', \csp 
    \delta \theta' = \half \frac {\pa \epsilon'} {\pa f} + a'
\eeq
one obtains $\srd = \sqrt{\pa f} \hat\srd$ with 
\beq    \label{eq:SoulutionNOneRedefinition}
    \hat\srd = \pm 1
\eeq
Here the $\sqrt{\pa f}$ factor merely reflects that $\theta$ is really a
conformal tensor of weight $-1/2$.

\paragraph{Reality condition}
If the $V^\dn$ is a real field, we expect the other components of the
same multiplet is also real in an irreducible multiplet.  This
necessitates a reality condition on $\theta$.  It can only be of the form
\beq
    \theta^* = \srl(z) \theta.
\eeq
which also has to be be invariant under \eqr{NOneActionOnSuperspace}.  
This leads to $\epsilon^* = z^{-2} \epsilon$ as before and also
\beq
    a^* = - \srl a\ .
\eeq
$\srl$ has to satisfy $\srl = \sqrt{-\pa g} \hat\srl = \hat\srl/z$ with:
$ \hat\srl = \pm 1$.  
Reality conditions are equivalent if and only if they are related by a
compatible reparameterization $(f,\srd)$ through
\beq
    \srl \to \srd^* \srl \srd^{-1}\ .
\eeq
It is easy to see that $f(z)= 1/z$ will flip the sign of $\hat\srl$.
Hence there is only one type of reality condition and one can set
$\hat\srl = 1$.

It also follows that the full reparameterization automorphism group,
$(f,\srd)$ that leaves $(g,\srl)$ invariant, consists of
conformal transformation plus conformal transformations combined with 
flipping the sign of odd elements.  The two correspond to the two
choices of sign for $\hat\srd$.  The ones with $\hat\srd = -1$ are
disconnected from the identity and not generated by $T$.  This choice
corresponds to an outer automorphism of the $N=1$ SCA.  A transition
function for the fiber of odd space over the even space has the same
choice.  The topological classes they allow are the spin
structures on the bosonic worldsheet.
The analysis of the $N=1$ SCA for the right movers is exactly the same.  


\paragraph{Boundary condition}
Now let us add a boundary to the worldsheet superspace parameterized by $z,
\theta$ and $\tilde z, \tilde \theta$.  The boundary must have a locus
on the even part of the worldsheet.  Again, it can be brought into
either positive real axis \eqr{PositiveRealAxis} or the unit circle
\eqr{UnitCircle}.  Requiring these equations to be invariant under the
superconformal transformations \eqr{NOneActionOnSuperspace} one finds
what infinitesimal transformation leaves the boundary invariant.  To
avoid repetition I will concentrate on the case of space-like boundary
at $\tilde z z = 1$.  The other cases are entirely analogous and easily
related to this one.

Substituting \eqr{NOneActionOnSuperspace} and its right mover 
counterparts in \eqr{UnitCircle}, one finds the conditions 
for the infinitesimal parameter on the boundary. In addition to 
\eqr{BoundaryConditionForCTSpacelike}, there is also 
\beq    \label{eq:NOneBoundaryConditionOna}
    \tilde z a \theta = z \tilde a \tilde \theta.
\eeq
If this has to hold for arbitrary $\theta$ and $\tilde \theta$, the only
solution is $a = \tilde a = 0$, i.e. this boundary breaks all
superconformal transformation.  The only boundary condition is
\eqr{BoundaryConditionForL}.  However, \eqr{NOneBoundaryConditionOna}
can have nontrivial solutions for $a$ if we consider a boundary which
does not span the whole odd part of the superworldsheet, but instead is
localized along a ``superline'' defined by 
\beq    \label{eq:GenericSuperBoundaryNIsOne}
    \tilde\theta = \sbr(\sigma)\theta\ .
\eeq
This is the most general conditions that can be imposed on the odd
part of the superworldsheet. \Eqr{NOneBoundaryConditionOna} then has
the solution
\beq
    \tilde z a + z \sbr \tilde a = 0\ .
\eeq
For consistency, one also has to require
\eqr{GenericSuperBoundaryNIsOne} to be invariant under such
supersymmetric transformation.  A straightforward calculation leads to 
the constraint $\sbr = \imath e^{-\imath\sigma} \hat\sbr$ with 
\beq    \label{eq:PossibleBoundayNIsOne}
    \hat\sbr = \pm 1\ .
\eeq
Again the $\imath e^{-\imath\sigma}$ phase reflects the tensor 
nature of $\theta$ and $\tilde\theta$.  The more important information 
lies in $\hat\sbr$.  At first sight, the two choices of signs seems
equivalent.  Under a reparameterization automorphism $(f, \srd)$ that
leave fixed the bosonic boundary, 
\beq
    \sbr \to \tilde\srd\sbr\srd^{-1}
\eeq
Hence one can flips the sign of $\hat\sbr$ by $\srd = -1$ with trivial
$f$, $\tilde f$ and $\tilde\srd$.  However, doing so would upset the
boundary conditions else where.  Consider a worldsheet with more than
one boundaries .  It is not possible to flips the sign of $\sbr$ for one
without affecting all the other.  This essentially has to do with the
above mentioned fact that reparameterization automorphisms with $\srd =
-1$ is disconnected from identity.  Therefore these two possibilities for
$\sbr$ correspond to two different types of boundaries.

Corresponding to them,
one has
\beq
    e^{-\sigma/2} a \pm \imath e^{-\sigma/2} \tilde a = 0.
\eeq
on the boundary, leading to the equation 
\beq
    G_m \pm \imath \tilde G_{-m} \BSk{} = 0
\eeq
This is the well-known boundary condition for the supercurrent used for
D-branes in Type II string theory.  The above analysis shows that it
corresponds to having a boundary in superspace defined by
\beq
    e^{\imath\sigma/2} \tilde\theta 
    = \pm\imath e^{-\imath\sigma/2} \theta 
\eeq
along with the more usual bosonic boundary.  It also shows how
limited the possibility is.  Any other space-like boundary one can
define is related to this one by a superconformal transformation.
For an annulus worldsheet with two boundaries, the relative choice of
signs on each translates to the choice of NS or R type of twisted $N=1$
algebra for open string theory and the possibility of inserting a
G-parity operator in the close string/boundary state picture.

\section {Boundary Conditions for $N=2$}

\paragraph{Multiplet structure}

Compared to $N=1$ SCA, $N=2$ SCA has two supercurrents $G^\pm$ instead of
one, and has an affine $U(1)$ current $J$ in addition.  Their sOPE's are 
given in the appendix.  From this and the Jacobi identity it is a
straightforward but tedious exercise to show that it has exactly three
families of irreducible tensor multiplets.
The more generic family is characterized by
two numbers, $h$, and $\lambda$ and consists of four fields, which we
denote as $V^\dn$, $V^-$, $V^+$, $V^\up$ respectively.  We shall call it
the \emph{long multiplet}.  Their sOPE's with the $N=2$ supercurrents
are
\beqar  \label {eq:OPEGLongMultiplet}
    G^+(z) \ldots V^\dn(w) &=& \frac {V^+ (w)} {z-w}\ , \nono
    G^-(z) \ldots V^\dn(w) &=& \frac {V^- (w)} {z-w}\ , \nono
    G^+(z) \ldots V^+(w) &=& G^-(z) \ldots V^-(w) = 0\ , \nono
    G^+(z) \ldots V^-(w) &=& \frac {h(1+\lambda)V^\dn (w)} {(z-w)^2} 
        + \frac {\half (1+\lambda) \pa V^\dn} {z-w} 
        + \frac {V^\up (w)} {z-w}\ , \nono
    G^-(z) \ldots V^+(w) &=& \frac {h(1-\lambda)V^\dn(w)} {(z-w)^2} 
        + \frac {\half (1-\lambda) \pa V^\dn(w)} {z-w} 
        - \frac {V^\up(w)} {z-w}\ , \nono
    G^+(z) \ldots V^\up(w) &=& 
        -\frac {(h+1/2)(1+\lambda)V^+(w)} {(z-w)^2} 
        - \frac {\half (1+\lambda) \pa V^+(w)} {z-w}\ , \nono
    G^-(z) \ldots V^\up(w) &=& 
        +\frac {(h+1/2)(1-\lambda)V^-(w)} {(z-w)^2} 
        + \frac {\half (1-\lambda) \pa V^- (w)} {z-w}\ , \nono
\eeqar
$V^\dn$, and $V^\pm$ are primary with respect to both
with (weight, charge) being $(h, 2h\lambda)$, $(h+1/2, 2h\lambda\pm
1)$ respectively.\footnote{
    $\Phi$ is primary with respect to a affine $U(1)$ current $J$ with
    charge $q$ if its sOPE with $J$ is 
    \[\label{eq:OPEJPrimary}
            J(z) \ldots \Phi(w)=  \frac {q \Phi(w)} {z-w} \ .
    \]}
As for $V^\up$, it is primary\footnote{
    \cite{Kiritsis:1987np} also contained explicit transformation
    properties of tensor components.  There the top component is
    \emph{not} primary with respect to $T$.  It is related to the
    convention here by $V^\up+(\lambda/2) \pa V^\dn$.}
with respect to $T$
with conformal weight $h+1$ but has the following OPE with $J$:
\beq    \label{eq:OPEJVup}
    J(z) \ldots V^\up(w) = 
    \frac {h (1-\lambda^2) V^\dn (w)} {(z-w)^2} 
    + \frac {2h\lambda V^\up (w)} {z-w}
\eeq
Note that $q= 2 h \lambda $ is the $U(1)$ charge of $V^\dn$ and what is
often called the charge of the whole multiplet.  
When $\lambda = \pm 1$ the
long multiplet becomes reducible 
into two shorter types of multiplets.
For our purpose the long multiplet is sufficient.

\paragraph{$N=2$ superworldsheet}

Without further ado let us turn a long multiplet into a superfield
\beq
    \msf V = V^\dn + \theta^+ V^- + \theta^- V^+ 
    + \theta^+ \theta^- V^\up
\eeq
by introducing two Grassmann variables $\theta^\pm$.
Now consider an infinitesimal $N=2$ superconformal transformation 
generated by 
\beq    \label{eq:GeneratorOfNTwoConformalTransformation}
    \frac 1 {2\pi\imath} \left(\myoI{dz} \epsilon (z) T(z) 
    + \myoI{dz} \xi(z) J(z) 
    + \myoI{dz} a^+(z) G^-(z) + \myoI{dz} a^-(z) G^+(z)\right).
\eeq
Here $\epsilon$ and $\xi$ are ordinary $\CC$ valued while $a^\pm$ are
Grassmann valued.  By performing
contour integration with the above OPE's, one find that
\beqar
    \delta \msf V &=& 
    \imath [\epsilon \pa + \half \pa \epsilon (\theta^+ \pa_{\theta^+} 
    + \theta^- \pa_{\theta^-}) 
    + h \pa \epsilon] V \\
    &+& \imath[\xi (- \theta^+ \pa_{\theta^+} + \theta^- \pa_{\theta^-}) 
    + 2 h \lambda \xi + h(1-\lambda^2) \pa \xi \theta^+ \theta^- ] V \nono
    &+& \imath[- \half (1 - \lambda) a^+ \theta^- \pa 
    + a^+ \pa_{\theta^+}
    + (1-\lambda) \pa a^+ (\half \theta^+ \theta^- \pa_{\theta_+}
        + h \theta^+)] V \nono
    &+& \imath[- \half (1 + \lambda) a^-  \theta^+ \pa
    + a^- \pa_{\theta^-}
    - (1+\lambda) \pa a^- (\half \theta^+ \theta^- \pa_{\theta_-} 
        + h \theta^-)] V \ . \nonumber
\eeqar

A novel aspect of $N=2$ SCA is that the derivation part of the variation
contain dependence on $\lambda$.  One can split it into two pieces as
$\lambda = \gamma + \breve\lambda$ with 
some arbitrary number $\gamma$, and treats it as an
universal piece for any tensor types.  It is clear that it should close
by itself in accordance with $N=2$ algebra for arbitrary $\gamma$.
Then the action of $N=2$ SCA on superspace is:
\beqar  \label{eq:NTwoConformalTransformationOnSupercoordinate}
    \delta z &=& \imath(\epsilon - \half (a^+ \theta^- + a^- \theta^+) 
    + \frac \gamma 2 ( a^+ \theta^- - a^- \theta^+))\ , \nono
    \delta \theta^\pm &=& \imath(\pm \xi \theta^\pm 
    + a^\pm \pm \half \pa a^\pm \theta^+\theta^- 
    - \half \gamma \pa a^\pm \theta^+ \theta^-)
\eeqar
We can write it more compactly by grouping $\theta^\pm$ into a
column vector $\theta$ and $a^\pm$ into $a$:
\beq
    a = \left(\bary {ll} a^+ \\ a^- \eary \right)\ , \csp 
    \theta = \left(\bary {ll} \theta^+ \\ \theta^- \eary \right)\ .
\eeq
Then \eqr{NTwoConformalTransformationOnSupercoordinate} becomes 
\beqar  \label{eq:NTwoSuperconformalTransformationCompact}
    \delta z &=& \imath [\epsilon + \frac {\theta^\top} 2 
        (\sigma^1 + \imath \gamma\sigma^2)a] \ , \nono
    \delta \theta &=& \imath[\half \pa \epsilon \theta 
    - \xi \sigma^3 \theta + a + \half \theta \theta^\top 
    (\sigma^1 + \imath \gamma\sigma^2) \pa a ]
\eeqar
The $N=2$ superworldsheet has a bigger odd part,
parameterized by $\theta^\pm$ and $\tilde\theta^\pm$.  It is fibered over
the even part. Note that so far $\gamma$
is an arbitrary parameter labelling $N=2$
superspace. Below it will be shown that different values of $\gamma$ are
in fact equivalent and can be set to zero.

\paragraph{Compatible reparameterization}

Along the same line as before, consider now reparameterization of the
superspace coordinate that is compatible with infinitesimal $N=2$ superconformal
transformations. Define 
\beq    \label{eq:ReparameterizationNTwo}
    z' = f(z) + \frac \omega 2 \theta^\top \imath \sigma^2 \theta, \csp 
    \theta' = \srd(z) \theta,
\eeq
where $\theta$ is a complex two-by-two matrix.  This is the most general
holomorphic reparametrization of superspace with ``bosonic''
parameters.  Requiring
\beqar
    \delta z' &=& \imath[\epsilon' 
    + \frac{\pa \epsilon'} {\pa f} 
        \frac \omega 2 \theta^\top \imath \sigma^2 \theta
    + \frac {{\theta'}^\top} 2 
        (\sigma^1 + \imath \gamma'\sigma^2)a'] \ , \nono
    \delta \theta' &=& \imath[\half \pa \epsilon' \theta' 
    - \xi' \sigma^3 \theta' 
    + a' 
    + \frac{\pa a'} {\pa f} 
        \frac \omega 2 \theta^\top \imath \sigma^2 \theta
    + \half \theta' {\theta'}^\top 
    (\sigma^1 + \imath \gamma'\sigma^2) \frac {\pa a'} {\pa f}]
\eeqar
leads to the following expression for the transformation of the parameters:
\beq
    \epsilon' = \pa f \epsilon\ , \csp a' = \srd a\ , \csp
    \xi' \sigma^3 = \xi \hat \srd \sigma^3 \hat \srd^{-1}  - \pa \hat \srd\,\hat\srd^{-1} \epsilon
\eeq
where $\hat \srd \equiv \srd / \sqrt{\pa f}$.  
There are a number of constraints on $\hat\srd$ and $\omega$, a minimal set of which are
\beq
    \pa \hat\omega = 0\ ,\csp 
    \sigma^3 \hat\srd \sigma^3 = \pm \hat\srd\ , \csp
    \hat \srd^{\top} (\sigma^1 + \imath \gamma'\sigma^2) \hat\srd 
    = (\sigma^1 + \imath (\gamma + \hat\omega) \sigma^2)\ ,
\eeq
where $\hat\omega = \omega / \pa f$. Corresponding to the choice of sign
in the last equation, there are two families of solutions:
\beq
    \hat \srd = e^{\phi \sigma^3}, \csp \gamma' = \gamma + \hat\omega\ ;
\eeq
and 
\beq
    \hat \srd = \sigma^1 e^{\phi \sigma^3}, \csp \gamma' = - \gamma - \hat\omega\ .
\eeq
Here $\phi$ is an arbitrary complex number. Using $\hat\omega$ we will 
set $\gamma$ to $0$ without loss of generality.  This choice fixes the 
$\omega$ parameter in to $0$ as well.

\paragraph{Reality condition}

To ensure the right number of degrees of freedom, one has to impose
\beq    \label{eq:NOneRealityCondition}
    \theta^* = \srl \theta
\eeq
where $\srl$ is a 2-by-2 matrix satisfying the consistency condition
\beq
    \srl^*\srl = 1
\eeq
In addition, \eqr{NOneRealityCondition} has to be compatible with the
conformal transformation \eqr{NTwoSuperconformalTransformationCompact}. 
This leads to
\beq
    a^* = - \srl a\ , \csp 
    \xi^* \sigma^3 = - \xi \hat\srl\sigma^3\hat\srl^{-1} 
    + \epsilon \pa\hat\beta\hat\beta^{-1}
\eeq
where $\hat\srl = \srl/\sqrt{-\pa g}$.  
$\hat\srl$ satisfies the following consistency conditions:
\beq
    \hat\srl^*\hat\srl = 1\ ,\csp 
    \sigma^3 \hat\srl \sigma^3 = \pm \hat\srl\ ,\csp
    \hat\srl^\top \sigma^1 \hat\srl 
    = \sigma^1\ .
\eeq
Corresponding to the choice of sign in the 2nd equation, there are two
families of solutions both parameterized by a real number $p$
\beq    \label{eq:RawRealityConditionNoncompactNIsTwo}
    \hat \srl = e^{\imath p \sigma^3}.
\eeq
and 
\beq    \label{eq:IntermediateSolutionRealityCompactNTwo}
    \hat \srl = \pm \sigma^1 e^{p \sigma^3}.
\eeq
The reality condition again transform under the reparameterization
$\srd$ as 
\beq    \label{eq:BetaTransformUnderXi}
    \srl \to \srd^* \srl \srd^{-1}\ .
\eeq
It is easy to see that one can always set $p$ to 0 (but see the caveat
below) and fix the sign in \eqr{IntermediateSolutionRealityCompactNTwo}.
The stabilizers
are
\beq    \label{eq:AutomorphismForNoncompactNTwo}
    \hat\srd = \pm e^{\phi \sigma^3} \csp \mbox{and} \csp
        \sigma^1 \pm e^{\phi \sigma^3}
\eeq
for $\hat\srl = \id$
and
\beq    \label{eq:AutomorphismForCompactNTwo}
    \hat\srd = e^{\imath \phi \sigma^3} \csp \mbox{and} \csp
        \sigma^1 e^{\imath \phi \sigma^3}
\eeq
for $\hat\srl = \pm \sigma^1$.

The meaning of the reparameterization automorphism is evident.  
$\srd$ is a function of $z$ taking value in a group $\cal G$.
For $\srl = 1$ $\cal G$ is $O(1,1)$ while for $\hat\srl = \sigma$ it
is $O(2)$. Ignore for now the topology of the worldsheet, the
group of reparameterization automorphisms has the same number of components
as $\cal G$.  $O(1,1)$ has four while $O(2)$ has two.  These components
correspond to the various solutions in
\eqr{AutomorphismForNoncompactNTwo} and \eqr{AutomorphismForCompactNTwo}
and in obvious ways.  They are the structure group of the Grassmann
fibers.  The global aspect of the compact case has been considered in
\cite{Cohn:1987wn}.

All things considered, there are two nonequivalent reality conditions. 
They correspond to different $N=2$ SCA's.  For $\hat\srl =
1$, the abelian group generated by $J_0$ is noncompact ---
this is known as the noncompact $N=2$ SCA.  It is easy to see that it cannot have unitary
highest weight representation.  On the other hand, $\srl = \sigma^1$
gives the conventional compact $N=2$ SCA with an $U(1)$ symmetry.  Most 
of the studies on $N=2$ SCA and SCFT have been on the compact version.

I should note here one caveat in the previous discussion.  When, as in
string theory, the worldsheet is compact in its spatial direction
$\sigma$, $\hat\srl$ in \eqr{RawRealityConditionNoncompactNIsTwo} can
have nontrivial winding as it takes value in $U(1)$.  This can still be
brought to the canonical form of $\hat\srl = 1$ by using $\hat\srd$ with
half as much winding.  However when the winding number of
$\hat\srl$ is odd, the necessary $\hat\srd$ is antiperiodic around
$\sigma$.  Its effect is to make $G^\pm$ also anti-periodic.  This gives
a twisted version of the $N=2$ mode algebra with half integer modes for
the supercurrents.  It is related to the $Z_2$ center of $O(1,1)$.

\paragraph{Boundary}
Now introduce a boundary to the $N=2$ superworldsheet.  Again concentrate
on the case of $z\bar z = 1$.  Invariance of this expression
under the $N=2$ superconformal transformation
\eqr{NTwoConformalTransformationOnSupercoordinate} leads to
\beq
    z^{-1} \theta^\top \sigma^1 a 
    + \tilde z^{-1} \tilde\theta^\top 
    \sigma^1 \tilde a = 0
\eeq
in addition to \eqr{BoundaryConditionForCTSpacelike}.  If the boundary
occupies the whole odd part of the superworldsheet, the only solution
is that $a = \tilde a = 0$, i.e. all supersymmetries are broken. In the
meantime there is no constraint on the abelian symmetries generated
by $J$ and $\tilde J$.
One can also limit the boundary to a subspace of the odd parts.
If a condition is imposed among $\theta$, the left moving $N=2$ SCA will
be broken down to a $N=1$ SCA.  Similarly one can do the same for the
right movers.  This returns us to boundary conditions of
$N=1$ SCA solved in the last section.  Here let us consider instead the
condition
\beq    \label{eq:ConditionOnFermionicParameterNIsTwo}
    \tilde \theta = \sbr(\sigma) \theta
\eeq
which can preserve a diagonal part of the two $N=2$ SCA's.

Requiring this equation to be invariant under \eqr{NTwoConformalTransformationOnSupercoordinate} leads to 
\beq    \label{eq:ConditionOnBosonicParameterNIsTwo}
    \tilde a = \sbr \epsilon, \csp 
    \tilde \xi \sigma^3 = 
    \xi \hat \sbr \sigma^3 \hat \sbr^{-1} 
    + \imath z^{-1} \epsilon \pa_\sigma \hat \sbr \hat \sbr^{-1}
\eeq
where $\hat \sbr = -\imath \exp{(\imath \sigma)}\sbr$.  
There are a number of constraints on $\hat\sbr$, 
a minimal set of which are
\beq
    \hat \sbr^{\top} \sigma^1 \hat\sbr 
    = \sigma^1,\csp
    \sigma^3 \hat\sbr \sigma^3 = \pm \hat\sbr\ .
\eeq
Corresponding to the choice of sign in the last equation, 
there are two families of solutions for $\hat\sbr$. It can either be 
$\exp{(\alpha\sigma^3)}$ or $\sigma^1 \exp{(\alpha\sigma^3)}$
Correspondingly $\tilde\gamma = \gamma$ and $\tilde\gamma = -\gamma$.
Hence the boundary types allowed depend on the $\gamma$ value of the 
superspace.  $\lambda = \tilde\lambda=0$ is the critical value allowing 
all possibilities. On top of that, the boundary has to be
compatible with the reality condition, which means
\beq
    \tilde \srl \sbr = \sbr^* \srl\ .
\eeq
From this we infer that $\alpha$ is real and imaginary respectively 
for noncompact and  compact $N=2$.

A reparameterization automorphism acts on the boundary conditions as 
\beq
    \sbr \to \tilde \srd \sbr \srd^{-1},
\eeq
which can be used to bring $\hat \sbr$ to $1$.  
However, as for the case of $N=1$, this does not mean there is only one
type of boundary.  Consider first the noncompact $N=2$.  The
solutions for $\sbr$ correspond to a map from the boundary to $O(1,1)$
and consists of four components:
\beq
    \hat \sbr = \pm e^{\phi \sigma^3}, \csp \mbox{or} \csp
     \pm \sigma^1 e^{\phi \sigma^3}.
\eeq
To bring them to $1$ one has to make use of the four
components of the reparameterization automorphism for noncompact $N=2$. 
Consider a worldsheet with more than one boundaries.  It is not hard to
see that $\sbr$ can be brought to $1$ on all boundaries if and only if
on all of them it lies in the same class.  Hence they really are 
four distinct types of boundaries.
For open string theory, the condition for the chiral fields can be deduced 
from \eqr{ConditionOnFermionicParameterNIsTwo} and
\eqr{ConditionOnBosonicParameterNIsTwo}.  
The four possibilities are grouped into two types.
The two boundary conditions of type B are \footnote
    {This nomenclature adheres to the convention in the literature
    \cite{Ooguri:1996ck}.} 
\beq    \label{eq:BoundaryFieldConditionOnChiralFieldsNIsTwoNoncompactB}
    T = \tilde T, \csp J = \tilde J, \csp 
    G^+ = \pm \tilde G^+, \csp G^- = \pm \tilde G^-
\eeq
and the two of type A are
\beq    \label{eq:BoundaryFieldConditionOnChiralFieldsNIsTwoNoncompactA}
    T = \tilde T, \csp J = \tilde J, \csp 
    G^+ = \pm \tilde G^+, \csp G^- = \pm \tilde G^-.
\eeq
Correspondingly there are four type of boundary state. They are
distinguished by their set of annihilators.  For type B they are
\beq    \label{eq:BoundaryStateConditionOnChiralFieldsNIsTwoNoncompactB}
    (L_m - \tilde L_{-m})\ , \csp (J_m + \tilde J_{-m})\ , \csp
    (G^+_m \pm \imath \tilde G^+_{-m})\ , \csp 
    (G^-_m \pm \imath \tilde G^-_{-m})\ .
\eeq
and for type A they are 
\beq    \label{eq:BoundaryStateConditionOnChiralFieldsNIsTwoNoncompactA}
    (L_m - \tilde L_{-m})\ , \csp (J_m - \tilde J_{-m})\ , \csp
    (G^+_m \pm \imath \tilde G^-_{-m})\ , \csp 
    (G^-_m \pm \imath \tilde G^+_{-m})\ .
\eeq

For compact $N=2$ SCA the situation is more subtle.  
The solutions for $\hat\sbr$ are
\beq
    \hat \sbr = \exp^{(\imath \alpha\sigma^3)} \csp \mbox{or} \csp
    \sigma^1 \exp^{(\imath\alpha\sigma^3)}\ , 
\eeq
It is a map from the boundary to $O(2)$.  
When the boundary is not a loop, say the time-like trajectory traced 
out by a propagating open string, there are two disjoint 
classes corresponding to the two components of $O(2)$, hence two type of
boundaries.  They are just the type A and B above.  Because a shift 
$\alpha \to \alpha + \pi$ changes the sign of all odd elements, the
subtypes within A or B are no longer distinct.  Algebraically this is
because the sign changing map is an outer automorphism for the
noncompact $N=2$ but an inner one for the compact $N=2$.  The two types 
are exactly what have been classified in
\cite{Ooguri:1996ck}.\footnote{
    Related boundary conditions were observed in the context of
    topological $\sigma$-models in an example of the A-type in
    \cite{Horava:1994ts} and considered for both types in
    \cite{Witten:1995fb}.}

However, when the boundary is a loop, for example corresponding to 
a boundary state, there are two infinite classes
of solutions parameterized by an integer $s$.  They corresponding to
a map from the boundary loop to the two components of $O(2)$ with
winding number $s$.  Therefore there are $2 \ZZ$ possible types of
boundary states.  
While the boundary conditions for zero winding are just those of type $A$ \eqr{BoundaryStateConditionOnChiralFieldsNIsTwoNoncompactA} or $B$ 
\eqr{BoundaryStateConditionOnChiralFieldsNIsTwoNoncompactB}
(with the subtypes no longer distinct),
the infinite series corresponding
nontrivial winding are not known before.  
A B-type boundary states $\BSk[B_s]$ with winding number $s$ are 
annihilated by 
\beqar \label {eq:BTypeS}
  (J_n &+& \tilde J_{-n} + s \frac c 3 \delta_{n,0}), \nono
  (L_n &-& \tilde L_{-n} - s \tilde J_{-n} 
        - s^2 \frac c 6 \delta_{n,0}), \csp \mbox{and} \nono
    (G^{\pm}_n &+& \imath \tilde G^{\pm}_{-n \pm s}).
\eeqar
An A-type boundary states $\BSk[A_s]$ with winding number $s$ are 
annihilated by
\beqar  \label {eq:ATypeS}
    (J_n &-& \tilde J_{-n} + s \frac c 3 \delta_{n,0}), \nono
    (L_n &-& \tilde L_{-n} + s \tilde J_{-n} 
        - s^2 \frac c 6 \delta_{n,0}), \csp \mbox{and} \nono
    (G^{\pm}_n &+& \imath \tilde G^{\mp}_{-n\mp s}).
\eeqar
Everything in these equations except the terms involving the central
charge are directly determined by substituting
\eqr{ConditionOnFermionicParameterNIsTwo} and
\eqr{ConditionOnBosonicParameterNIsTwo} in
\eqr{GeneratorOfNTwoConformalTransformation} and its right moving
counterpart with $\hat\sbr = \exp{(\imath s\sigma \sigma^3)}$ for type B
and $ \hat\sbr=\sigma^1 \exp{(\imath s\sigma \sigma^3)}$ for type A.  The
result is consistent with the $N=2$ algebra without central extension. 
The central terms are obtained by demanding consistency with the
full algebra, in particular $\{G^+_m, G^-_{-m}\}$.

Now that these states are found, it is natural to look for their
algebraic interpretation.  The original type A and B boundary 
conditions were classified \cite{Ooguri:1996ck} by the
$Z_2$ outer automorphism of $N=2$ SCA which inverts the sign of the $U(1)$
current:
\beq    \label {eq:Conjugation}
        L \to L, \csp J \to -J, \csp G^\pm \to G^\mp \ .
\eeq
These two new series should also correspond to families of outer
automorphism.  In fact there is a well-known family of transformations
of $N=2$ SCA parameterized by a real number $s$.  It is known as
spectral flow and changes the boundary condition of $G^\pm$ by
$\exp{(2\pi\imath s)}$ \cite{Schwimmer:1987mf}:
\beqar  \label {eq:SpectralFlow}
        J_n &\to& J_n + s \frac c 3 \delta_{n,0}\ ; \nono
        L_n &\to& L_n + s J_n + s^2 \frac c 6 \delta_{n,0}\ ; \nono
        G^{\pm}_n &\to& G^{\pm}_{n\pm s} \ .
\eeqar
It was obtained by the adjoint action of an affine-$U(1)$ transformation
with monodromy $\exp{(2\pi\imath s)}$.  The case of $s =\pm 1/2$ is
important for understanding spacetime supersymmetry in $N=2$
compactifications of superstring. When $s$ is integral, it does not
change the periodicity condition of $G^\pm$ so becomes an automorphism
of the $N=2$ SCA.  It is in fact an outer automorphism because the
affine-$U(1)$ transformation leading to it corresponds to an element of
the loop group of $U(1)$ disconnected from the identity.\footnote
    {Recently the related spectral flow in and nontrivial elements of
    the loop group of $SL(2,\RR)$ has found application in a different
    capacity \cite{Maldacena:2000hw}.}
It is easy to
see that the $B_s$ series of boundary states are related to this
automorphism.  Now, by following \eqr{SpectralFlow} with a conjugation
automorphism, one obtains another outer automorphism:
\beqar    \label {eq:ConjugateSpectralFlow}
        G^{\pm}_n &\to& G^{\mp}_{n\mp s} \ ; \nono
        J_n &\to& -J_n - s \frac c 3 \delta_{n,0} \ ; \nono
        L_n &\to& L_n + s J_n + s^2 \frac c 6 \delta_{n,0} \ .
\eeqar
which evidently corresponds to the $A_s$ series of boundary states.

This view of the nontriviality of the boundary conditions fit nicely
with the above derivation of these boundary conditions from geometry of
the superworldsheet.  $s$ is the winding numbers of $\sbr$ in the loop
group of $O(2)$.  If one insists on unwinding it by say $\Xi$ without
introduce winding at other boundary components, one has to introduce
either a Chern class for the $U(1)$ bundle or, as a singular limit, the
insertion of integral spectral flow operators.
It is also worth noting that just as mirror symmetry relates
type A and B boundary conditions, so it also relates $A_s$ and $B_{-s}$.

\section {Free field realization}

Here I give a free field realization of the two new infinite series of
boundary states in \eqr{BTypeS} and \eqr{ATypeS}, in terms of $d$ pairs
of complex bosons and complex fermions with a Hermitian metric $(\cal
G_{i\bar j})^* = \cal G_{\bar i j} \equiv \cal G_{j \bar i}$:
\beq
    \brak{\alpha^i_m,\alpha^{\bar j}_n} = 
        m{\cal G}^{i\bar j}\delta_{m+n}\ , \csp 
    \brac{\psi^i_m,\psi^{\bar j}_n} = {\cal G}^{i\bar j}\delta_{m+n}\ , \csp
    i, \bar j = 1\ldots d.
\eeq
and the same for the right movers.  It is sufficient to consider the
sector with half integral moding for all the fermions.
The compact $N=2$ SCA is realized as (with normal ordering)
\beqar
    G^+_n = \sum_{p+q=n}\psi^i_p\alpha^{\bar i}_q\ , &\csp& 
    G^-_n = \sum_{p+q=n}\psi^{\bar i}_p\alpha^i_q\ , \nono
    J_n = \sum_{p+q=n}\psi^i_p\psi^{\bar i}_q\ ,&\csp&
    L_n = \sum_{p+q=n}\alpha^i_p\alpha^{\bar i}_q 
        + \frac {q-p} 2 \psi^i_p\psi^{\bar i}_q\ .
\eeqar
It is easy to verify that \eqr{BTypeS} is satisfied provided that 
$\BSk{B_s}$ is annihilated by\footnote{
    In the language of D-branes, the $R$ matrix here for type B and
    later for type A encodes the geometric and gauge field
    configurations of a D-branes, as defined in \cite{Ooguri:1996ck}. 
    It also has to satisfy certain consistency condition (unitary or
    anti-unitarity)as discussed \emph{ibid}.}
\beqar
    (\alpha^i_m + {R^i}_j \tilde\alpha^j_{-m}), &\csp&
    (\alpha^{\bar i}_m + {R^{\bar i}}_{\bar j} \tilde\alpha^{\bar j}_{-m}), \nono
    (\psi^i_m +\imath  {R^i}_j \tilde\psi^j_{-m+s}), &\csp&
    (\psi^{\bar i}_m +\imath  {R^{\bar i}}_{\bar j} \tilde\psi^{\bar j}_{-m-s}),
\eeqar
where $({R^i}_j)^* = {R^{\bar i}}_{\bar j}$ is unitary. Their kernel is
$2^{|s|}$ dimensional.  For $s\geq 0$, one solution can be constructed
from the $SL(2,\RR)\times SL(2,\RR)$ vacuum $\CSk[\emptyset]$ as
\beqar
        \BSk[B_{-s}]_{0} &=& \exp[-\sum_{m>0} 
          \frac 1 m (R_{i \bar j} a^i_{-m} \tilde a^{\bar j}_{-m}
                + R_{\bar i j} a^{\bar j}_{-m} \tilde a^i_{-m})] \nono
        &&\exp [-\sum_{m>0} \imath (
                R_{i \bar j} \psi^i_{-m} \tilde \psi^{\bar j}_{-m-s} +
                R_{\bar i j} \psi^{\bar i}_{-m-s} \tilde \psi^j_{-m})] 
                        \nono
        &&\prod_{k>0}^{s} 
              (\psi^{\bar i}_{k-s} + \imath {R^{\bar i}}_{\bar j} 
                        \tilde \psi^{\bar j}_{-k}) 
        \CSk[\emptyset] \ .
\eeqar
Here $R_{\bar i j}=\cal G_{\bar i i}{R^i}_j$, and $R_{i \bar j}=\cal
G_{i \bar i} {R^{\bar i}}_{\bar j}$.  A complete set of basis can be
formed by $\BSk[B_s]_{0}$ and applying products of $(\psi^{\bar i}_{k-s}
- \imath {R^{\bar i}}_{\bar j} \tilde \psi^{\bar j}_{-k})$ on it for
$0<k<s$.

For $s\leq 0$ one solution is 
\beqar
        \BSk[B_s]_{0} &=& \exp[-\sum_{m>0} 
          \frac 1 m (R_{i \bar j} a^i_{-m} \tilde a^{\bar j}_{-m} 
                 +R_{\bar i j} a^{\bar j}_{-m} \tilde a^i_{-m})] \nono
        &&\exp [-\imath\sum_{m>0} (
                R_{i \bar j} \psi^i_{-m+s} \tilde \psi^{\bar j}_{-m} +
                R_{\bar i j} \psi^{\bar i}_{-m} \tilde \psi^j_{-m+s})] 
                        \nono
        &&\prod_{k>0}^{-s} (\psi^i_{-k} + \imath {R^i}_j \tilde \psi^j_{k+s}) 
        \CSk[\emptyset]\ .
\eeqar
Applying $(\psi^i_{-k} - \imath {R^i}_j \tilde \psi^j_{k+s})$ on it for
$0<k<-s$ and one can span the whole solution spaces.

$\BSk[A_s]$ can be obtained similarly.  The solitons lies in the common
kernel of
\beqar
    (\alpha^i_m + {R^i}_{\bar j} \tilde\alpha^{\bar j}_{-m}), &\csp&
    (\alpha^{\bar i}_m + {R^{\bar i}}_{j} \tilde\alpha^{j}_{-m}), \nono
    (\psi^i_m +\imath  {R^i}_{\bar j} \tilde\psi^{\bar j}_{-m-s}), &\csp&
    (\psi^{\bar i}_m +\imath  {R^{\bar i}}_j \tilde\psi^{j}_{-m+s}),
\eeqar
which is again $2^{|s|}$ dimensional.  Here $({R^i}_{\bar j})^* =
{R^{\bar i}}_{j}$ is anti-unitary.\footnote
    {By that I mean ${R^i}_{\bar i} \cal G_{i \bar j} {R^{\bar j}}_j =
    \cal G_{j\bar i}$.}
For $s\geq 0$ one solution is
\beqar
        \BSk[A_{s}]_{0} &=& \exp[-\sum_{m>0} \frac 1 m (
           R_{i j} a^i_{-m} \tilde a^{j}_{-m} +
           R_{\bar i \bar j} a^{\bar j}_{-m} \tilde a^{\bar i}_{-m})] \nono
        &&\exp [-\sum_{m>0} \imath (
           R_{i j} \psi^i_{-m-s} \tilde \psi^{j}_{-m} +
           R_{\bar i \bar j} \psi^{\bar i}_{-m} 
                \tilde \psi^{\bar j}_{-m-s})] \nono
        &&\prod_{k>0}^s (\psi^i_{-k} + \imath {R^i}_{\bar j} 
                \tilde \psi^{\bar j}_{k-s}) 
        \CSk[\emptyset] \ .
\eeqar
Applying products of $(\psi^i_{-k} - \imath {R^i}_{\bar j} \tilde
\psi^{\bar j}_{k-s})$ on $\BSk[A_{s}]_{0}$, for $0<k<s$ to get a
basis spanning the whole kernel.

For $s\leq 0$ one has
\beqar
        \BSk[A_{s}]_{0} &=& \exp[-\sum_{m>0} \frac 1 m (
                R_{i j} a^i_{-m} \tilde a^{j}_{-m} +
                R_{\bar i \bar j} a^{\bar j}_{-m} \tilde a^{\bar i}_{-m})] \nono
        &&\exp [-\sum_{m>0} \imath (
                R_{i j} \psi^i_{-m} \tilde \psi^{j}_{-m+s} +
                R_{\bar i \bar j} \psi^{\bar i}_{-m+s} \tilde \psi^{\bar j}_{-m})] 
                        \nono
        &&\prod_{k>0}^{-s} 
                (\psi^{\bar i}_{k+s} + \imath {R^{\bar i}}_{j} \tilde 
                \psi^{j}_{-k}) 
        \CSk[\emptyset] \ .
\eeqar
Applying $(\psi^{\bar i}_{k+s} - \imath {R^{\bar i}}_{j} \tilde
\psi^{j}_{-k})$ on $\BSk[A_{s}]_{0}$ for $0<k<-s$ to span the whole
solution space.

\vspace{2ex}

It would be interesting to study these new types of boundary states in
the context of general $N=2$ superconformal field theories.  Another
natural extension of the analysis in this paper would be to the
cases of $N=3$ and $N=4$ SCA's.

\vspace{2ex}

I would like to thank V.~Kac for illuminating replies to my question
about outer automorphisms, L.~Alvarez-Gaume, I.~Brunner, M.B.~Halpern,
C.~Helfgott, and Y.~Oz for discussions.

\section*{Appendix: Summary of $N=0,1,2$ SCA OPE's}

The Lie algebra of the mode operators of two chiral field $\phi$ and
$\psi$ are determined by the singular part of their operator product
expansion (sOPE), which is denoted in this paper by $\phi(z) \ldots
\phi(w)$.

\subsection*{$N=0$}
\beq    \label{eq:TTOPE}
    T(z)\ldots T(w) = \frac {c/2} {(z-w)^4} + \frac {T(w)} {(z-w)^2} 
    + \frac {T(w)} {z-w}\ .
\eeq

\subsection*{$N=1$}
$G$ is primary with conformal weight $3/2$.  Its sOPE with itself is
\beq    \label{eq:GGOPE}
	G(z)\ldots G(w) =\frac {2c/3} {(z-w)^3} +  \frac {2T(w)} {z-w}\ .
\eeq

\subsection*{$N=2$}
$G^\pm$ and $J$ are primary with 
conformal weight $3/2$ and $1$.  The rest of their sOPEs are
\beqar    \label {eq:NTwoOPE}
        G^+(z)\ldots G^+(w) &=& G^-(z)G^-(w) \,=\,~ 0 \ ,\nono
        G^+(z)\ldots G^-(w) &=& \frac {c/3} {(z-w)^3}
        + \frac {J(w)} {(z-w)^2} + \frac {\pa J(w) /2} {z-w} +
          \frac {T(w)} {z-w} \ ,\nono
        J(z)\ldots G^{\pm}(w) &=&\, \pm \frac{G^{\pm}(w)}{z-w} \ ,\nono
        J(z)\ldots J(w) &=& \frac {c/3} {(z-w)^2} \ .
\eeqar

\bibliographystyle{hepunsrt}

\bibliography{string-literature}

\end {document}